\documentclass[a4paper,12pt]{article}
\pdfoutput=1 

\usepackage{jheppub} 

\usepackage[T1]{fontenc} 

\title{Michel parameters in radiative muon decay}

\author[a,b,1]{A.B. Arbuzov,\note{Corresponding author.}}

\author[b]{T.V. Kopylova}

\affiliation[a]{Bogoliubov Laboratory for Theoretical Physics, JINR, 
Dubna 141980, Russia}
\affiliation[b]{Dubna State University, Dubna 141982, Russia}

\emailAdd{arbuzov@theor.jinr.ru}

\abstract{
Radiative muon and tau lepton decays are described within the model-independent 
approach with the help of generalized Michel parameters.
The exact dependence on charged lepton masses is taken into account.
The results are relevant for modern and future experiments on muon and 
tau lepton decays.
}


\begin{document} 
\maketitle
\flushbottom

\section{Introduction}

Studies of muon decay are one of the keystones in particle physics.
This decay is almost a pure weak-interaction process. That allows us to use
it for high-precision tests of the Standard Model (SM) and searches for
new physics, see e.g. the review~\cite{Kuno:1999jp}. 

According to the principle of lepton universality that is adapted in the SM, 
the pure leptonic modes of tau lepton decays are described by the same Feynman
diagrams as the corresponding muon decays. The only difference is coming from
the fact that the masses of charged leptons are changed. 
As concerning models beyond the SM, some of them predict violation of lepton 
universality for tau leptons, assuming
that the third generation is more strongly coupled to some ``new physics'' than
the first two ones~\cite{Pich:1995vj,Rouge:2000um}. 
This motivates experimental searches for new physics
in tau decays at various high-energy machines including the
LHC, see e.g.~\cite{Celis:2014asa,Hayreter:2015cia}. 
In particular, a new method to probe magnetic and electric dipole moments of tau 
lepton using precise measurements of the differential rates of radiative leptonic decays 
at high-luminosity $B$-factories was proposed in~\cite{Eidelman:2016aih}.
Note that a high precision has been already reached in measurements of 
radiative leptonic tau decays at the $B$-factories, see e.g.~\cite{Lees:2015gea}. 

The precision of the Michel parameters~\cite{Michel:1949qe,Kinoshita:1957zz,Kinoshita:1957zza} 
definition from 
tau decays~\cite{Albrecht:1990zj,Ackerstaff:1998yk,Ammar:1996xh,Heister:2001me,Abdesselam:2014uea}
is not yet competing with the one achieved in muon decays. 
Nevertheless, high statistics of tau lepton observation at 
the $B$-factories Belle and BaBar, and at the LHC provides 
good perspectives for studies of various tau lepton decay modes.

\section{Preliminaries and notation}

Within the SM, muon decays are described by interactions of vector currents
formed by left fermions. Meanwhile, many models beyond the SM predict contributions
of other kinds. Since the energy scale of new physics is (most likely) higher
than the electroweak scale, the corresponding contributions can be parameterized
by contact four-fermion interactions with different currents and coupling constants.
The matrix element of the muon decay can be presented 
in the general form~\cite{Fetscher:1986uj}
\begin{eqnarray}
\label{M_element}
{\cal M} = 4\frac{G_0}{\sqrt{2}}
\sum_{\stackrel{\gamma=\mathrm{S,V,T}}{\epsilon,\omega=\mathrm{R,L}}}
g_{\epsilon\omega}^{\gamma} \,
\langle \, \bar{l}_\epsilon|\Gamma^\gamma|\nu_l \rangle \,
\langle \bar{\nu}_\tau |\Gamma_\gamma|\tau_\omega \rangle ,
\end{eqnarray}
see for details the Particle Data Group review~\cite{Agashe:2014kda}   
and references therein.
Here index $\gamma$ denotes the type of the interaction: scalar (S), 
vector (V) or tensor (T); and $\Gamma^{\gamma}$ are $4\times 4$ matrices defined in
terms of the Dirac matrices:
\begin{equation}
\Gamma^{\rm S} = 1, \qquad \Gamma^{\rm V} = \gamma^\mu, \qquad
\Gamma^{\rm T} = \frac{1}{\sqrt{2}}\,\sigma^{\mu\nu}
= \frac{i}{2\sqrt{2}} (\gamma^\mu\gamma^\nu - \gamma^\nu\gamma^\mu).
\end{equation}
The indices $\omega$ and $\epsilon$ denote the chiralities of the
initial and final charged leptons, respectively. For a
given pair ($\omega$, $\epsilon$) the chiralities of neutrinos are
uniquely determined. Tensor interactions can contribute only for opposite
chiralities of the charged leptons. This leads to the existence of 
10 complex coupling constants, $g_{\epsilon\omega}^{\gamma}$.
The Standard Model predicts $g_{LL}^{V}=1$ and all others being
zero. Choosing the arbitrary phase by defining $g_{LL}^{V}$ to be real 
and positive leaves 19 real numbers to be determined by the experiment.  
As long as one is interested in the relative strengths of the couplings,
it is convenient to require the following normalization condition:
\begin{eqnarray} \label{norm}
N &\equiv& \frac{1}{4} \left( {|g_{LL}^{S}|}^2 + {|g_{LR}^{S} |}^2
  + {|g_{RL}^{S}|}^2 + {|g_{RR}^{S} |}^2 \right)
\nonumber   \\
&+& \left( {|g_{LL}^{V}|}^2 + {|g_{LR}^{V} |}^2
 + {|g_{RL}^{V}|}^2 + {|g_{RR}^{V} |}^2 \right)
 + 3  \left( {|g_{LR}^{T}|}^2 + {|g_{RL}^{T} |}^2 \right) = 1.
\end{eqnarray}
This restricts the allowed ranges of the coupling constants to
$|g^{S} |\leq 2$, $|g^{V} |\leq 1$, and $|g^{T} |\leq \frac{1}{\sqrt{3}}$. 
The overall normalization can be incorporated into $G_0$ which then
accounts for deviations from the Fermi coupling constant $G_{\mathrm{Fermi}}$. 
We have to note that 
the precision of the muon life time measurement is much higher than the one
obtained for definition of $G_{\mathrm{Fermi}}$ via other parameters of the SM, 
i.e. $\alpha_\mathrm{QED}$, $M_Z$, $M_W$ etc. Unfortunately, for 
this reason the extremely precise measurement of $G_{\mathrm{Fermi}}$ does not 
provide any valuable test of models beyond the SM.

In this paper we provide a parameterization of radiative muon (tau) decays
in terms of the generalized Michel parameters~\cite{Kinoshita:1957zz,Kinoshita:1957zza}
which are certain bilinear combinations of the coupling constants $g_{\epsilon\omega}^{\gamma}$. 
Our aim is to take into account the exact dependence on the charged fermion masses.

\section{Radiative muon decay}

At the Born level the decay 
\begin{eqnarray}
\mu^-(p_\mu) \to e^-(p_e)\ +\ \nu_\mu(k_1)\ +\ \bar{\nu}_e(k_2)\ +\ \gamma(p_\gamma)
\end{eqnarray}
was considered by many authors since late 
fifties~\cite{Kinoshita:1957zz,Fronsdal:1959zzb,Eckstein:1959:xxx}.
Here we follow the notation adapted in the review~\cite{Kuno:1999jp} and
represent the differential width of this decay in the form
\begin{eqnarray} \label{rad_decay_diff}
&& \frac{d \Gamma(\mu^{\pm}\to e^{\pm}\bar{\nu}\nu\gamma)}{dx\,dy\,d\Omega_e\,d\Omega_\gamma}
= \Gamma_0 \frac{\alpha_\mathrm{QED}}{64\pi^3}\frac{\beta_e}{y} \biggl[ F(x,y,d)
\mp \beta_e P_\mu \cos\theta_e G(x,y,d) 
\nonumber \\
&& \qquad \mp P_\mu \cos\theta_\gamma H(x,y,d) \biggr],
\nonumber \\
&& \Gamma_0 = \frac{G_{\mathrm{Fermi}}^2m_\mu^5}{192\pi^3}, \qquad 
d = 1-\beta_e\cos\theta_{e\gamma}, \qquad \beta_e = \sqrt{1-\frac{m_e^2}{E_e^2}}\, ,
\end{eqnarray}
where $\Omega_{e,\gamma}$ are solid angles of the observable final state particles; 
$\theta_e$ and $\theta_\gamma$ are the angles between the muon spin and the electron
and photon momenta, respectively; $P_\mu$ is the muon polarization degree;
$\theta_{e\gamma}$ is the angle between the electron and photon momenta;
$x$ and $y$ are energy fractions of electron and photon, respectively,
$x\equiv 2E_e/m_\mu$ and $y\equiv 2E_\gamma/m_\mu$.

Functions $F$, $G$, and $H$ are polynomials in the electron to muon mass ratio:
\begin{eqnarray}
\mathcal{F}(x,y,d) = \sum_{k=1}^{5} \left(\frac{m_e}{m_\mu}\right)^k\mathcal{F}^{(k)},
\qquad \mathcal{F} \equiv F,\ G,\ H.
\end{eqnarray}
We computed these functions in the tree-level approximation.
Analytical calculations were performed with the help of the {\tt FORM} computer
language~\cite{Vermaseren:2000nd}.
The explicit expression for these functions are given in Appendix~\ref{App_A} below.
These functions depend on the generalized Michel parameters $\rho$, $\eta$, $\bar\eta$,
$\xi$, $\delta$, $\kappa$, $\alpha$, and $\beta$, see Appendix~\ref{App_B}. 
The dependence on $\bar\eta$,  $\kappa$, $\alpha$, and $\beta$ provides important
additional information about the structure of weak interactions with respect to
the studies of non-radiative muon and tau decays, see~\cite{Eichenberger:1984gi}. 
Note that the dependence on $\alpha$, and $\beta$ appear only in terms suppressed 
by the first power of the mass ratio $r\equiv m_e/m_\mu \approx 5\cdot 10^{-3}$  
(or $m_\mu/m_\tau \approx 6\cdot 10^{-2}$ for the $\tau\to\mu\nu\bar{\nu}$ decay). 
The presence of the contributions proportional to the first power
of the mass ratio (and in general in odd powers) is a non trivial effect. 
Presumably, it is related to a hidden spin flip. An analogous effect takes 
place in the one-loop corrections to polarized muon decay spectrum: 
expansion of the exact result~\cite{Arbuzov:2001ui} for
these corrections contains some terms which are linear in the electron 
to muon mass ratio.

The contributions suppressed by the electron to muon mass ratio
can be important for modern high-precision experiments on muon
and especially tau lepton radiative decays. As one can see from~(\ref{F_i}), 
(\ref{G_i}), and (\ref{H_i}), the most important higher order effect is given
by function $F^{(1)}(x,y,d)$ since it is the only one being linear in the
charged lepton mass ratio. Moreover, this function depends on the generalized
Michel parameters $\alpha$ and $\beta$ which do not appear in the lowest
approximation in $r$. In addition, this function contains terms proportional to $d^{-1}$
which provide a considerable enhancement in the kinematical domain of collinear
radiation of photons along the electron direction of motion. Thus one can hope
that future experiments on radiative muon and tau decays can provide
additional information on the Michel parameter values. 

We verified our results by comparison with ones the existing in the literature.
We have checked that in the case of $V-A$ interactions our results completely 
agree with the formulae given in~\cite{Kuno:1999jp}. For the case of general
interactions in the limit $m_e \ll m_\mu$, we performed a comparison with the results
of~\cite{Fronsdal:1959zzb}. Here the agreement is not complete.
Namely, we have the coincidence for functions $F(x,y,d)|_{m_e\to0}$ and $G(x,y,d)|_{m_e\to0}$.
But we encountered a small discrepancy for $H(x,y,d)|_{m_e\to0}$, which
can be removed after the following modification in one term in the Appendix 
of~\cite{Fronsdal:1959zzb}:
\begin{eqnarray}
F_1^{T\gamma} = -2[x^2y^2(1-y)+2x^3y] \to F_1^{T\gamma} = -2[x^2y^2(1-y)+2x^3y^2].
\end{eqnarray}
Moreover, one more obvious misprint in~\cite{Fronsdal:1959zzb} has to be 
corrected in eq.~(4b):
\begin{eqnarray}
[\Delta + (2/\mu^2x^2)] \to [\Delta + (2m_e^2/\mu^2x^2)].
\end{eqnarray}
Note that the expression in square brackets above is a certain approximation 
of quantity $d$ given in~(\ref{rad_decay_diff}). 

The NLO QED radiative corrections to this process 
(for the case of the pure $V-A$ interactions)
were recently calculated in~\cite{Fael:2015gua} with taking into account the
exact dependence on the final state charged lepton mass.
Earlier, the one-loop corrections to radiative muon decay were
also considered in~\cite{Fischer:1994pn,Arbuzov:2004wr}.
Besides the QED effects, radiative corrections to the $W$ boson propagator
within the Standard Model become numerically relevant~\cite{Fael:2013pja}. 
The processes of five-body leptonic decays of muons and tau leptons
can be also used to get an additional information about the
structure of weak interactions, see e.g.~\cite{Flores-Tlalpa:2015vga}.

\section{Conclusions}

Our results can be used in the analysis of high-precision experimental 
data on $\tau$ and muon radiative decays in order to extract the 
generalized Michel parameters. 
The exact dependence on the final charged lepton mass is taken into account.
The corresponding effect for certain cases is linear in the ratio of the
final and initial charged lepton masses. So it becomes numerically important
first of all for studies of the $\tau\to \mu\bar{\nu}\nu\gamma$ decay.
A correction to the result of paper~\cite{Fronsdal:1959zzb} is also made. 

\acknowledgments

We are grateful to D.~Epifanov and S.~Eidelman for statement
of the problem and useful discussions and to E.~Akzhigitova for
the help with cross checks of the results. 

\appendix

\section{Functions $F$, $G$, and $H$
\label{App_A}}

Here we give the explicit formulae for functions $F$, $G$, and $H$
which appear in~(\ref{rad_decay_diff}):
\begin{eqnarray} \label{F_i}
F^{(0)} &=&  32\rho \biggl( \frac{16x^2y + 12xy^2 + 4 y^3 + 8x^3- 3y^2 - 6 x^2 - 6xy}{3 d}
  - \frac{10}{3}x^2y^2 - \frac{8x^3y}{3} - 2xy^3
\nonumber \\
  &-& \frac{xy^2}{3} - \frac{5x^2y}{3} - \frac{4 x^3}{3} 
  + xy    + x^2 
+ xyd\bigl( \frac{4x^2}{3} + x^2y+xy^2 +\frac{3}{4}xy  - \frac{x}{2} \bigr)
    - \frac{x^3y^2d^2}{6}(2 + y) \biggr)
\nonumber \\
 &+& \bar{\eta} 8xy^2\biggl( 2y + 4x - xd - xyd - 2 \biggr)
  + 48\frac{y^2+2xy+2x^2-y^3-2x^3-3xy^2-4x^2y}{d} 
\nonumber \\
  &+& 96x^3y + 112x^2y^2 + 56xy^3  + 48x^2(x+y-1) + 16xy^2 
  - 48xy 
\nonumber \\
  &+& 4xyd\bigl( 6x - 7xy(1+y) - 12x^2 - 9x^2y\bigr) + 6x^3y^2d^2(2 + y),
\nonumber \\
F^{(1)} &=& \beta \biggl( \frac{24x(x+y-1)}{d} - \frac{4y^2}{d} 
  + 4xy^2 - 12x(xy+x+y-1)
  - x^2y^2d + 6x^2yd \biggr)
\nonumber \\
 &+& \alpha \biggl( \frac{12x(1-x-y)}{d}
  + 6x(xy + x + y - 1) - 3x^2yd \biggr),
\nonumber \\
F^{(2)} &=&  16 \rho \biggl( \frac{8(x+y)(3-4x-4y)}{3xd^2} 
  + \frac{18y^2-12x^2+20xy-16x-12y}{3d} 
\nonumber \\
 &+& 2x^2y  - \frac{2xy^2}{3} + 2x^2 + 2xy 
  + \frac{8x}{3} - \frac{x^2yd(2+y)}{2} \biggr)
  + 16\bar{\eta} \biggl(  - xy^2 - \frac{2y^2}{d} \biggr)
\nonumber \\
 &+& \frac{192(x^2+y^2+2xy-x-y)}{x d^2} 
  + \frac{96x^2-112y^2-96xy+96y}{d}
  - 48x^2y + 16xy^2 
\nonumber \\
 &-& 48x^2  - 48xy  + 12x^2yd(2+y),
\nonumber \\
F^{(3)} &=& 96 \eta \biggl( 4\frac{x+y-1}{xd^2} + \frac{2(x-y)}{d} 
  - x \biggr),
\nonumber \\
F^{(4)} &=&  64\rho \frac{8+6x+6y-3xyd}{3x d^2}
  + 96\frac{xyd-2y-2x}{x d^2},
\nonumber \\
F^{(5)} &=& \eta \biggl(  - \frac{384}{x d^2} \biggr),
\end{eqnarray}
\begin{eqnarray} 
G^{(0)} &=& 8\xi \biggl(  \frac{4x^3+6x^2y-4x^2-2xy}{d} + \frac{4xy^2}{3d} 
  - 3x^3y - \frac{5}{3}x^2y^2 - 2x^3 - x^2y + 2x^2 
\nonumber \\
 &+& \frac{x^3yd(2+y)}{2} \biggr)
 + 32\delta\xi \biggl( \frac{xy}{d} - \frac{10xy^2}{9d} + \frac{2x^2}{d}
  - \frac{4x^2y}{d} - \frac{8x^3}{3d} - x^2 
   + \frac{5x^2y}{6} 
\nonumber \\
 &+& \frac{11x^2y^2}{9}  + \frac{4x^3}{3} + 2x^3y 
 - \frac{x^3yd}{3}(2+y) \biggr)
 + 8\kappa\xi xy^2\biggl( x - \frac{2}{d} \biggr),
\nonumber
\end{eqnarray}
\begin{eqnarray} \label{G_i}
G^{(2)} &=& 8\xi \biggl( \frac{8(1-x-y)}{d^2} 
  + \frac{2xy}{d} - \frac{4x^2}{d} + 2x^2 + x^2y \biggr)
\nonumber \\
 &+& 16\delta\xi \biggl( \frac{8(4x+4y-3)}{3d^2} 
  - \frac{2xy}{d} + \frac{20x^2}{3d} - \frac{10x^2}{3} - \frac{5x^2y}{3} \biggr),
\nonumber \\
G^{(4)} &=& \xi  \frac{64}{d^2} - \delta\xi \frac{640}{3d^2}\, ,
\nonumber \\
G^{(1)} &=& 0, \qquad G^{(3)} = 0, \qquad G^{(5)} = 0,
\end{eqnarray}
\begin{eqnarray} \label{H_i}
H^{0} &=&
  8\xi \biggl( \frac{2y(x^2+y^2-y-x)}{d} + \frac{14xy^2}{3d} 
  - \frac{7}{3}xy^3 - x^3y + \frac{2xy^2}{3}
  - 3x^2y^2  + x^2y 
\nonumber \\
 &+& \frac{7}{6}x^2y^3d + x^3y^2d - \frac{x^2y^2d}{2} 
  - \frac{x^3y^3d^2}{4} \biggr)
  + 8\delta\xi \biggl( \frac{4y(3x+3y-4x^2-4y^2)}{3d} - \frac{104xy^2}{9 d} 
\nonumber \\
 &+& \frac{40}{9}xy^3 + \frac{8x^3y}{3} + \frac{20}{3}x^2y^2 + \frac{4xy^2}{9}
  - 2 x^2y + x^2y^2d - \frac{20}{9}x^2y^3d 
  - \frac{8}{3}x^3y^2d + \frac{2}{3}x^3y^3d^2 \biggr)
\nonumber \\
  &+& 8\kappa\xi  \biggl(\frac{2xy^2}{d} + 2xy^3 + 3x^2y^2 - 4xy^2 - x^2y^3d \biggr),
\nonumber \\
H^{(2)} &=&
  8\xi \biggl( \frac{8y(1-y-x)}{x d^2} 
     + \frac{14y^2}{3d} + \frac{2xy}{d}  - \frac{4y}{d}
 - \frac{2xy^2}{3}  + x^2y - \frac{x^2y^2d}{2} \biggr)
\nonumber \\
 &+& 8\delta\xi \biggl( \frac{16y(4y+4x-3)}{3xd^2} 
  - \frac{116y^2}{9d} - \frac{4xy}{d} + \frac{8y}{d}
  + \frac{20xy^2}{9} - \frac{10x^2y}{3} + \frac{5x^2y^2d}{3} \biggr)
\nonumber \\
       &+& 16\kappa\xi \biggl(  - xy^2 - \frac{2y^2}{d} \biggr),
\nonumber \\
H^{(4)} &=&
    32\xi  \biggl( \frac{2y}{x d^2} - \frac{y}{d} \biggr) 
  + 320\delta\xi  \biggl( \frac{y}{3d} - \frac{2y}{3 x d^2} \biggr),
\nonumber \\
H^{(1)} &=& 0, \qquad H^{(3)} = 0, \qquad H^{(5)} = 0.
\end{eqnarray}

\section{Generalized Michel parameters
\label{App_B}}

For the sake of completeness, we give here the explicit expressions for the generalized
Michel parameters defined by Kinoshita and Sirlin~\cite{Kinoshita:1957zz,Kinoshita:1957zza}, 
see also~\cite{Agashe:2014kda} for details, via the coupling constants 
$g_{\epsilon\omega}^{\gamma}$ which enter~(\ref{M_element}):
\begin{eqnarray}
\rho &=& \frac{3b + 6c}{16} = \frac{3}{4} + \frac{3}{4}\cdot\frac{(2c-a)}{16},
\qquad
\eta = \frac{\alpha - 2\beta}{16}\, ,
\qquad
\bar{\eta} = \frac{a + 2c}{16}\, ,
\nonumber \\ 
\kappa\xi &=& \frac{a_1+2c_1}{16}\, ,
\qquad
\delta\xi =  \frac{6c_1-3b_1}{16}\, ,
\qquad
\xi =  \frac{14c_1-3a_1-4b_1}{16}\, ,
\nonumber
\end{eqnarray}
\nonumber \\
\begin{eqnarray}
\alpha &=& 8\mathrm{Re}\biggl( g^V_{RL}(g^{S*}_{LR}+6g^{T*}_{LR}) 
  + g^V_{LR}(g^{S*}_{RL}+6g^{T*}_{RL}) \biggr),
\qquad
\beta = - 4\mathrm{Re}( g^V_{RR}g^{S*}_{LL} + g^V_{LL}g^{S*}_{RR} ),
\nonumber \\ 
a &=& 16(|g^V_{RL}|^2 + |g^V_{LR}|^2) + |g^S_{RL}+6g^T_{RL}|^2 
    + |g^S_{LR}+6g^T_{LR}|^2,
\nonumber \\
b &=& 4(|g^V_{RR}|^2 + |g^V_{LL}|^2) + |g^S_{RR}|^2 + |g^S_{LL}|^2,
\nonumber \\
c &=& \frac{1}{2}|g^S_{RL} - 2g^T_{RL}|^2 + \frac{1}{2}|g^S_{LR} - 2g^T_{LR}|^2,
\nonumber \\
a_1 &=& 16(|g^V_{RL}|^2 - |g^V_{LR}|^2) + |g^S_{RL}+6g^T_{RL}|^2 
     - |g^S_{LR}+6g^T_{LR}|^2,
\nonumber \\
b_1 &=& 4(|g^V_{RR}|^2 - |g^V_{LL}|^2) + |g^S_{RR}|^2 - |g^S_{LL}|^2,
\nonumber \\
c_1 &=& \frac{1}{2}|g^S_{RL} - 2g^T_{RL}|^2 - \frac{1}{2}|g^S_{LR} - 2g^T_{LR}|^2.
\end{eqnarray}
Note that we used above the normalization condition~(\ref{norm}), which
can be re-written also as
\begin{eqnarray}
a + 4b + 6c = 16.
\end{eqnarray}
In the limit of the pure $V-A$ interactions we have $\rho=\delta=\frac{3}{4}$, $\xi=1$,
$\eta=\bar\eta=\kappa=\alpha=\beta=0$.


\begin{thebibliography}{99}

\bibitem{Kuno:1999jp}
  Y.~Kuno and Y.~Okada,
  \emph{Muon decay and physics beyond the standard model},
  \emph{Rev. Mod. Phys.} {\bf 73} (2001) 151
  [hep-ph/9909265].

\bibitem{Pich:1995vj}
  A.~Pich and J.~P.~Silva,
  \emph{Constraining new interactions with leptonic $\tau$ decays},
  \emph{Phys. Rev.} {\bf D 52} (1995) 4006
  [hep-ph/9505327].

\bibitem{Rouge:2000um}
  A.~Rouge,
  \emph{Tau lepton Michel parameters and new physics},
  \emph{Eur. Phys. J.} {\bf C 18} (2001) 491
  [hep-ph/0010005].

\bibitem{Celis:2014asa}
  A.~Celis, V.~Cirigliano and E.~Passemar,
  \emph{Model-discriminating power of lepton flavor violating $\tau$ decays},
  \emph{Phys. Rev.} {\bf D 89} (2014) 9,  095014
  [arXiv:1403.5781].

\bibitem{Hayreter:2015cia}
  A.~Hayreter and G.~Valencia,
  \emph{Spin correlations and new physics in τ-lepton decays at the LHC},
  \emph{JHEP} {\bf 1507} (2015) 174
  [arXiv:1505.02176].

\bibitem{Eidelman:2016aih}
  S.~Eidelman, D.~Epifanov, M.~Fael, L.~Mercolli, and M.~Passera,
  \emph{$\tau$ dipole moments via radiative leptonic $\tau$ decays},
  \emph{JHEP} {\bf 1603} (2016) 140
  [arXiv:1601.07987 [hep-ph]].

\bibitem{Lees:2015gea}
  BaBar Collaboration, J.~P.~Lees et al., 
  \emph{Measurement of the branching fractions of the radiative leptonic $\tau$ decays $\tau \to e\gamma\nu\bar{\nu}$ and $\tau \to \mu\gamma\nu\bar{\nu}$ at $B{\small A}B{\small AR}$},
  \emph{Phys. Rev.} {\bf D 91} (2015) 051103
  [arXiv:1502.01784].

\bibitem{Michel:1949qe}
  L.~Michel,
  \emph{Interaction between four half spin particles and the decay of the $\mu$ meson},
  \emph{Proc. Phys. Soc.} {\bf A 63} (1950) 514.

\bibitem{Kinoshita:1957zz}
  T.~Kinoshita and A.~Sirlin,
  \emph{Muon Decay with Parity Nonconserving Interactions and Radiative Corrections in the Two-Component Theory},
  \emph{Phys. Rev.} {\bf 107} (1957) 593.

\bibitem{Kinoshita:1957zza}
  T.~Kinoshita and A.~Sirlin,
  \emph{Polarization of Electrons in Muon Decay with General Parity-Nonconserving Interactions},
  \emph{Phys. Rev.} {\bf 108} (1957) 844.

\bibitem{Albrecht:1990zj}
  ARGUS Collaboration, H.~Albrecht et al.,
  \emph{Determination of the Michel parameter in tau decay},
  \emph{Phys. Lett.} {\bf B 246} (1990) 278.

\bibitem{Ackerstaff:1998yk}
  OPAL Collaboration, K.~Ackerstaff et al., 
  \emph{Measurement of the Michel parameters in leptonic tau decays},
  \emph{Eur. Phys. J.} {\bf C 8} (1999) 3
  [hep-ex/9808016].

\bibitem{Ammar:1996xh}
  CLEO Collaboration, R.~Ammar et al., 
  \emph{A Measurement of the Michel parameters in leptonic decays of the tau},
  \emph{Phys. Rev. Lett.}  {\bf 78} (1997) 4686.

\bibitem{Heister:2001me}
  ALEPH Collaboration, A.~Heister et al., 
  \emph{Measurement of the Michel parameters and the nu/tau helicity in tau lepton decays},
  \emph{Eur. Phys. J.} {\bf C 22} (2001) 217.

\bibitem{Abdesselam:2014uea}
  Belle Collaboration, A.~Abdesselam et al., 
  \emph{Study of Michel parameters in leptonic $\tau$ decays at Belle},
  arXiv:1409.4969.

\bibitem{Fetscher:1986uj}
  W.~Fetscher, H.~J.~Gerber and K.~F.~Johnson,
  \emph{Muon Decay: Complete Determination of the Interaction and Comparison with 
        the Standard Model},
  \emph{Phys. Lett.} {\bf B 173} (1986) 102.

\bibitem{Agashe:2014kda}
  Particle Data Group Collaboration, K.~A.~Olive et al.,
  \emph{Review of Particle Physics},
  \emph{Chin. Phys.} {\bf C 38} (2014) 090001.

\bibitem{Fronsdal:1959zzb}
  C.~Fronsdal and H.~Uberall,
  \emph{mu-Meson Decay with Inner Bremsstrahlung},
  \emph{Phys. Rev.}  {\bf 113} (1959) 654.

\bibitem{Eckstein:1959:xxx}
  S.G.~Eckstein, R.H.~Pratt,
  \emph{Radiative muon decay},
  \emph{Annals of Physics} {\bf 8} (1959) 297.

\bibitem{Vermaseren:2000nd}
  J.A.M.~Vermaseren,
  \emph{New features of FORM},
  math-ph/0010025.

\bibitem{Eichenberger:1984gi}
  W.~Eichenberger, R.~Engfer and A.~Van Der Schaaf,
  \emph{Measurement Of The Parameter Eta-bar In The Radiative Decay Of The Muon As A Test Of The V-a Structure Of The Weak Interaction},
  \emph{Nucl. Phys.} {\bf A 412} (1984) 523.

\bibitem{Arbuzov:2001ui}
  A.~B.~Arbuzov,
  \emph{First order radiative corrections to polarized muon decay spectrum},
  \emph{Phys. Lett.} {\bf B 524} (2002) 99
  [hep-ph/0110047].

\bibitem{Fael:2015gua}
  M.~Fael, L.~Mercolli and M.~Passera,
  \emph{Radiative $\mu$ and $\tau$ leptonic decays at NLO},
  \emph{JHEP} {\bf 1507} (2015) 153
  [arXiv:1506.03416].

\bibitem{Fischer:1994pn}
  A.~Fischer, T.~Kurosu and F.~Savatier,
  \emph{QED one loop correction to radiative muon decay},
  \emph{Phys. Rev.} {\bf D 49} (1994) 3426.

\bibitem{Arbuzov:2004wr}
  A.~B.~Arbuzov and E.~S.~Scherbakova,
  \emph{One loop corrections to radiative muon decay},
  \emph{Phys. Lett.} {\bf B 597} (2004) 285
  [hep-ph/0404094].

\bibitem{Fael:2013pja}
  M.~Fael, L.~Mercolli and M.~Passera,
  \emph{W-propagator corrections to $\mu$ and $\tau$ leptonic decays},
  \emph{Phys. Rev.} {\bf D 88} (2013)  093011
  [arXiv:1310.1081 [hep-ph]].

\bibitem{Flores-Tlalpa:2015vga}
  A.~Flores-Tlalpa, G.~Lopez Castro and P.~Roig,
  \emph{Five-body leptonic decays of muon and tau leptons},
  \emph{JHEP} {\bf 1604} (2016) 185
  [arXiv:1508.01822 [hep-ph]].

\end{thebibliography}
\end{document}